\documentclass[fleqn,usenatbib]{mnras}

\usepackage[dvipsnames]{xcolor}
\usepackage{newtxtext,newtxmath}
\usepackage[T1]{fontenc}
\usepackage{ae,aecompl}
\usepackage{comment}
\usepackage[all]{hypcap} 
\usepackage{multirow}
\usepackage{marginnote}
\usepackage{subfigure}
\usepackage{graphicx}    % Including figure files
\usepackage{amsmath}    % Advanced maths commands

\usepackage{bm}
\usepackage{mathrsfs}
\title[Gradient measurement of polarization and applications]{Gradient measurement of synchrotron polarization diagnostic: Application to spatially separated emission and Faraday rotation regions}

\author[Wang et al.]{Ru-Yue Wang$^1$, Jian-Fu Zhang$^2$, Alex Lazarian$^3$, Hua-Ping Xiao$^2$ and Fu-Yuan Xiang$^2$\\
$^1$Department of Physics, Xiangtan University, Xiangtan, Hunan 411105, China; hpxiao@xtu.edu.cn; jfzhang@xtu.edu.cn; \\
$^2$Key Laboratory of Stars and Interstellar Medium, Xiangtan University, Xiangtan, Hunan 411105, China;\\
$^3$Astronomy Department, University of Wisconsin, Madison, WI 53711, USA
}

\date{Accepted XXX. Received YYY; in original form ZZZ}
\pubyear{2021}

\begin{document}
\label{firstpage}
\pagerange{\pageref{firstpage}--\pageref{lastpage}}
\maketitle

\begin{abstract}
Considering the spatially separated polarization radiation and Faraday rotation regions to simulate complex interstellar media, we study synchrotron polarization gradient techniques' measurement capabilities. We explore how to trace the direction of projected magnetic field of emitting-source region at the multi-frequency bands, using the gradient technique compared with the traditional polarization vector method. Furthermore, we study how Faraday rotation density in the foreground region, i.e., a product of electron number density and parallel component of magnetic fields along the line of sight, affects the measurement of projected magnetic field. Numerical results show that synchrotron polarization gradient technique could successfully trace projected magnetic field within emitting-source region independent of radio frequency. Accordingly, the gradient technique can measure the magnetic field properties for a complex astrophysical environment.
\end{abstract}


\section{Introduction}

Turbulence and magnetic field are widespread in astrophysical environments (Armstrong, Rickett \& Spangler~\citeyear{ARS1995}; Elmegreen \& Scalo~\citeyear{ES04}; Chepurnov \& Lazarian~\citeyear{CheLa2010}; Haverkorn~\citeyear{Haverkorn}), and turbulent motions of fluid result in magnetic field fluctuations. The turbulent magnetic field plays a vital role in many key astrophysical processes, such as star formation (see McKee \& Ostriker~\citeyear{MckeOstr2007}; Mac Low \& Klessen~\citeyear{MacKle2004} ), propagation and acceleration of cosmic rays (Jokipii \citeyear{Jokipii1966}; Schlickeiser~\citeyear{SCH2002}; Yan \& Lazarian~\citeyear{YanAlex2008}), magnetic dynamo action (Fyfe, Joyce \& Montgomery~\citeyear{Fyfe1977}; Malyshkin \& Boldyrev~\citeyear{Malyshkin2009}; Xu \& Lazarian \citeyear{XuLazarian2016}), density structures in the Interstellar Medium (Higdon~\citeyear{Higdon1996}; Xu, Ji \& Lazarian~\citeyear{XuJiLazarian2019}), heat conduction in galaxy clusters (Bu, Wu \& Yuan~\citeyear{Bu2016}; Yuan et al.~\citeyear{Yuan2015}) and turbulent reconnection (Lazarian \& Vishniac~\citeyear{LV99}, hereafter LV99; Kowal et al.~\citeyear{Kowal2009}; Eyink, Lazarian \& Vishniac~ \citeyear{Eyink2011}). In addition, the properties of magnetic field in diffuse Galactic media are essential for resolving key cosmological problems related to the detection of cosmic microwave background (CMB) polarization arising from the enigmatic B-modes of cosmological origin (Cho \& Lazarian~\citeyear{ChoLazarian2010}; Planck Collaboration et al.~\citeyear{Planck2016a}). 

It is universally acknowledged that magnetic turbulence properties is extremely difficult to measure in astrophysical environments. For instance, the effect of using traditional Faraday rotation of polarized synchrotron emission is a significant impediment for studying emission of atomic hydrogen at high redshifts (Cho, Lazarian \& Timbie~\citeyear{Cho2012}). In general, the in-situ observational information is obtained from the solar wind turbulence, providing an important reference for understanding the turbulence within the ISM and galaxies. In fact, the challenge of studying MHD turbulence by using observations is inevitably line-of-sight integrated. Therefore, the purpose of technique development of magnetic field measurement is to extract turbulence information from the integrated observation information. In view of this, we will explore in this paper the new way of studying magnetic field using synchrotron gradients technique (SGT) first proposed in Lazarian et al. (\citeyear{L2017}) and Lazarian \& Yuen (\citeyear{L2018a}), further explored and elaborated in Zhang et al. (\citeyear{Zhang19a}, \citeyear{Zhang19b}).

The synchrotron studies are based on the fact that the relativistic electrons spiraling through the magnetic field produce synchrotron radiation that could reveal the magnetic field information (Waelkens, Schekochihin \& En{\ss}lin~\citeyear{Waelkens2009}; Junklewitz \& En{\ss}lin~\citeyear{Junklewitz}; Lazarian \& Pogosyan~\citeyear{LP12}, hereafter LP12). Considering statistics of synchrotron radiation intensity, one can obtain the properties of magnetic fields perpendicular to the line of sight (LOS). It is well known that one of the main characteristics of synchrotron radiation is its polarization effect. However, the polarization emission signal inevitably encounters Faraday rotation depolarization effect in the process of propagation, resulting in the direction variation of intrinsic polarization vector. Although this effect distorts inherent polarization signal, it provides a way to understand the properties of turbulence volume that radiative signal passes, such as the component of magnetic fields along the LOS and electron distribution.

With the purpose of understanding the anisotropy and compressibility of MHD turbulence, LP12 provided a theoretical description of synchrotron intensity fluctuations arising from magnetic turbulence. This study using synchrotron emission has opened avenues for quantitative studies of magnetic turbulence in the Galactic and extragalactic ISM. Some of the analytical descriptions presented by LP12 have been successfully testified by MHD turbulence simulations (Herron et al. ~\citeyear{Herron16}). Moreover, theoretical expressions of synchrotron polarization were proposed in Lazarian \& Pogosyan~(\citeyear{LP16}, hereafter LP16), where they introduced several measurement methods of turbulence to obtain power spectral slopes and correlation scales of underlying magnetic turbulence. The analytical expressions for one-point and two-point statistical techniques provided by LP16 have been confirmed by synthetic simulations (Lee, Lazarian \& Cho ~\citeyear{Lee2016}; Zhang et al. ~\citeyear{Zhang16},~\citeyear{ZhangL18}). Since polarized synchrotron intensity fluctuations are anisotropic -- with a stronger correlation along the direction of mean magnetic field, the ratio of structure function in two different directions can be used to trace the direction of mean magnetic field. The theoretical prediction proposed in LP12 in terms of quadrupole ratio modulus of synchrotron intensity was generalized to the case of synchrotron polarization intensity (Lee, Cho \& Lazarian~\citeyear{Lee2019}; Wang, Zhang \& Xiang~\citeyear{Wang2020}) in order to quantitatively measure anisotropy of MHD turbulence. Using the quadrupole ratio modulus to study compressible MHD turbulence, Wang, Zhang \& Xiang~(\citeyear{Wang2020}) revealed the anisotropic properties of Alfv{\'e}n, slow and fast modes, in good agreement with earlier direct numerical results (Cho \& Lazarian ~\citeyear{Cho2003}).

Synchrotron polarization gradients were first used to constrain the sonic Mach number of interstellar turbulence (Gaensler et al.~\citeyear{Gaensler2011}; Burkhart, Lazarian \& Gaensler~\citeyear{Burkhart2012}). Furthermore,
synchrotron intensity gradients (SIGs) and synchrotron polarization gradients (SPGs) were identified as a means for tracing magnetic field (Lazarian et al.~\citeyear{L2017}; Lazarian \& Yuen~\citeyear{L2018a}). The theoretical justification of the gradients is based on 
fundamental properties of magnetic turbulence (Goldreich \& Sridhar~\citeyear{GS95}, henceforth GS95; see a monograph by Beresnyak \& Lazarian \citeyear{Beresnyak2019} for the latest development in the field), magnetic reconnection (LV99; see Lazarian et al. \citeyear{Lazarian2020a} for a recent review)
and the theory of synchrotron fluctuations (LP12 and LP16). The utility of SIGs and SPGs for probing magnetic field was successfully tested with numerical simulations and confirmed by comparison with observational data, e.g. Planck polarization data. The ability of SPGs to recover 3D distribution of magnetic field was demonstrated in Lazarian \& Yuen~(\citeyear{L2018a}). This new technique has also been applied to tracing projected mean magnetic field direction in super-Alfv{\'e}nic turbulence regime in terms of multifrequency measurement (Zhang et al.~\citeyear{Zhang19a}). As a result, SPGs and SIGs become strong counterparts for another technique for magnetic field studies, i.e., velocity gradient technique (Gonz{\'a}lez-Casanova \& Lazarian~\citeyear{Casanova2017}; Yuen \& Lazarian~\citeyear{Yuen2017}; Lazarian \& Yuen~\citeyear {L2018b}). The synergies of these three techniques shed new light on studying magnetic fields in multiphase media of Milky Way and external galaxies.\footnote{The gradient techniques are uniquely suitable for the use with interferometers. It was first shown in Lazarian et al. (\citeyear{L2017}) that the full magnetic field structure can be restored with only high spacial frequencies measured by the interferometer (see Lazarian, Yuen \& Pogosyan~ \citeyear{Lazarian2020b} for the theoretical justification). This paves the way for studying magnetic fields using interferometers without adding single dish data.}

Various measures can be constructed with the gradients of synchrotron polarization, a few of which were considered in Herron et al.~(\citeyear{Herron18a, Herron18b}). However, they missed the point of the ability of these measures to trace magnetic field directions. Zhang, Liu \& Lazarian~(\citeyear{Zhang19b}) explored the flexibility of these polarization measures to robustly predict the direction of Galactic projected magnetic fields, and identified the spatial gradient from the maximum of the radial component of the polarization directional derivative which can trace the direction of magnetic field with the highest accuracy. Another new technique, synchrotron polarization derivative with respect to the squared wavelength $dP/d\lambda^2$, has been proposed to measure the local magnetic field (Lazarian \& Yuen~\citeyear{L2018a};  Zhang et al.~\citeyear{Zhang2020}), the recovering of which is critical to measure the 3D magnetic field of Milky Way.

Until now, the development and application of synchrotron gradient techniques are only in spatially coincident synchrotron emission and Faraday rotation regions. In reality, synchrotron polarization radiation obtained by the observers could be from spatially separated regions in a real astrophysical environment. The prospect of synchrotron gradient techniques for studying magnetic fields in the ISM motivates us to carry out the study in the case of spatially separated synchrotron emission and Faraday rotation regions. We want to explore whether synchrotron polarization gradient is still an excellent tool to measure the direction of magnetic field for more complex astrophysical situation.

The structure of this paper is organized as follows. In Section \ref{sec:foundation}, we provide descriptions regarding theoretical foundation of MHD turbulence, synchrotron polarization radiation, diagnostics of synchrotron polarization, expectations for the observed signal and gradient measurement technique. Section \ref{sec:data} describes the procedure of numerical simulation of MHD turbulence. The numerical results are presented in Section \ref{sec:results}. Finally, the implications of our findings are discussed and concluded in Section \ref{sec:discussions}.

\section{Theoretical foundation}
\label{sec:foundation}
\subsection{Fundamental theory of MHD turbulence}
\label{sec:theory}
The modern understanding of MHD turbulence theory dated from the Goldreich \& Sridhar (GS95) pioneering work. The additional developments, e.g., the introduction of the concept of the local system of reference 
(LV99; Cho \& Vishniac ~\citeyear{Cho2000}; Maron \& Goldreich ~\citeyear{Maron2001}), generalization from trans-Alfv{\'e}nic to arbitrary Alfv{\'e}n Mach numbers (LV99; Galtier et al. ~\citeyear{Galtier2000}), and generalization from incompressible turbulence to compressible one (Lithwick \& Goldreich ~\citeyear{Lithwick2001}; 
Cho \& Lazarian ~\citeyear{ChoL2002}, ~\citeyear{Cho2003}; Kowal \& Lazarian ~\citeyear{Kowal2010}), were realized within the framework established within the
aforementioned pioneering study.

GS95 proposed the scale-dependent anisotropy of incompressible MHD turbulence, i.e., the smaller the turbulence scale, the more elongated the anisotropic structure of eddies. It should be noticed that the theory was formulated in the global system of reference, where GS95 relations are in fact invalid. This issue was corrected in subsequent works where the notion of local system of reference was introduced and successfully tested (LV99; Cho \& Vishniac~\citeyear{Cho2000}; Maron \& Goldreich ~\citeyear{Maron2001}; Cho, Lazarian \& Vishniac~\citeyear{Cho2002}). The critical balance concept of MHD turbulence theory should be formulated in the local reference system, i.e., the local magnetic field surrounding the eddies. Specifically, the LV99 work provided a comprehensive interpretation of GS95 theory from the perspective of turbulence eddy, where magnetic reconnection occurs within one eddy turnover time and the motion of eddies perpendicular to the magnetic field is not influenced by magnetic tension. In this description, the importance of local system of reference is self-evident. The eddies interact only with the magnetic field in their vicinity and their rotations are aligned with the local magnetic field.

The magnetization of the media is described by the Alfv{\'e}nic Mach number, i.e., $M_{\rm A}=V_{\rm L}/V_{\rm A}$, where $V_{\rm L}$ represents the injection velocity of turbulence driving at the scale $L_{\rm inj}$, and $V_{\rm A}$ is the Alfv{\'e}nic velocity determined by magnetic field $B$ and plasma density $\rho $. Considering incompressible MHD turbulence for $M_{\rm A}\sim 1$, GS95 found that the relationship of turbulence anisotropy between parallel and perpendicular directions of local magnetic field can be written as 
\begin{equation}
l_{\|}\propto l_{\perp}^{2/3}, \label{eq:aniso}
\end{equation}
which was derived under the condition of critical balance of $v_{l}l_{\perp}^{-1}= V_{\rm A}l_{\|} ^{-1}$, where $l_{\|}$  and  $l_{\perp}$ are the parallel and perpendicular scales of the eddy, respectively, and $v_{l}$ is the velocity at the scale $l$. 

Additionally, GS95's anisotropy descriptions are generalized to $M_{\rm A}<1$ (LV99) and $M_{\rm A}>1$ (Lazarian~\citeyear{L2006}), respectively. The $M_{\rm A}<1$ case, i.e., sub-Alfv{\'e}nic turbulence, shows a weak turbulence from the driving scale $L_{\rm inj}$ to the transition scale $l_{\rm trans}=L_{\rm inj}M_{\rm A}^2$, while a strong turbulence occurs from $l_{\rm trans}$ to the dissipation scale $l_{\rm diss}$. Over the inertial range of [$l_{\rm diss}$, $l_{\rm trans}$], the relationship of the parallel and perpendicular scale of eddies is described as
\begin{equation}
l_{\|}\approx L_{\rm inj}^{1/3}l_{\perp}^{2/3}M_{\rm A}^{-4/3}. \label{anis}
\end{equation}
which gets back to the predictions of GS95 theory for $M_{\rm A}\sim1$. The turbulent velocity is expressed by
\begin{equation}
v_{\perp}=V_{\rm A}(\frac{l_{\perp}}{L_{\rm inj}})^{1/3}M_{\rm A}^{4/3}=V_{L}(\frac{l_{\perp}}{L_{\rm inj}})^{1/3}M_{\rm A}^{1/3},
\end{equation}
which shows Kolmogorov-type ($v_{\perp}\propto l_{\perp}^{1/3}$) cascade perpendicular to local magnetic field. Here, the velocity gradient scale is calculated by $v_{\perp}/l_{\perp}\approx l_{\perp}^{-2/3}$, that is, the largest velocity gradient corresponds to the smallest eddy. This relation is also conducive to the study of magnetic field gradient, because magnetic field and velocity are symmetric in Alfv{\'e}nic turbulence.

For super-Alfv{\'e}nic turbulence ($M_{\rm A}>1$), the motions of turbulence are marginally constrained by magnetic field when the turbulence scale is larger than the transition scale $l_A=L_{\rm inj}M_{\rm A}^{-3}$, so the turbulence shows an essentially hydrodynamic Kolmogorov property. Since there is no information about the gradient of magnetic field, the direction of magnetic field cannot be measured by the gradient method in this range. When the scale is smaller than $l_A$, magnetic field becomes again important. The scaling is identical to trans-Alfv{\'e}nic MHD turbulence theory if we identify the effective injection scale with $l_A$. In this case, we expect the direction of gradient of magnetic field is perpendicular to the magnetic field. When gradient technique is used to trace the magnetic field for trans-Alfv{\'e}nic turbulence, large-scale structure is better to be removed.\footnote{This removal does not affect the directions measured by gradients as theory of gradients in Lazarian, Yuen \& Pogosyan~ (\citeyear{Lazarian2020b}) demonstrates that the largest spatial frequencies contain all the necessary information for tracing magnetic fields with gradients. Therefore, filtering out the low spatial frequencies does not degrade the ability of studying magnetic fields with gradients. This was empirically demonstrated in Lazarian et al. (\citeyear{L2017}) and Lazarian \& Yuen (\citeyear{L2018b}).}

Although the GS95 theory is also confronting with a different voice, the development of gradient techniques would not be affected because the gradient measurements independent of scaling slope are constrained in the inertial range of turbulence cascade. Some modifications to GS95 theory at the driving or dissipation scale do not affect the present gradient techniques tracing magnetic field. For instance, several studies (Boldyrev~\citeyear{Boldyrev2005},~\citeyear{Boldyrev2006}; Mason, Cattaneo \& Boldyrev~\citeyear{Mason2006}) attempted to explain the numerical simulations in Maron \& Goldreich (\citeyear{Maron2001}) and proposed that a particular process termed as dynamical alignment can modify the GS95 spectrum from $k^{-5/3}$ to $k^{-3/2}$. However, further research showed that the deviations from the GS95 slope have the transient nature, i.e., localized in the vicinity of the injection scale, and do not proceed through the entire inertial range (Beresnyak \& Lazarian~\citeyear{Beresnyak2010}; Beresnyak~\citeyear{Beresnyak2013}; Beresnyak~\citeyear{Beresnyak2014}). This explains why the initial low resolution numerical simulations were 
producing a more shallow slope while the higher resolution studies obtained $k^{-5/3}$ in agreement with the GS95 predictions. It is worth mentioning that the anisotropy expected in Bolryrev’s modification of MHD turbulence theory is 
clearly inconsistent with numerical simulations (see more in Beresnyak \& Lazarian ~\citeyear{Beresnyak2019}; Beresnyak ~\citeyear{Beresnyak2019a}). 

More recently, a modification of MHD theory for motions at scales close to the dissipation scale was proposed (see Mallet et al. ~\citeyear{Mallet2019}). The
testing of the corresponding predictions requires computational abilities that far exceed those available now or in the 
near future. There are also serious conceptual problems with the “reconnection mediated MHD turbulence”,
as magnetic reconnection is a part and parcel of the turbulent cascade (LV99; Eyink et al. ~\citeyear{Eyink2011}, ~\citeyear{Eyink2013}; 
Lazarian et al. ~\citeyear{Lazarian2020a}). In a word, the attempts to revise the GS95 theory have not been able to change the paradigm so far. 

\subsection{Synchrotron polarization radiation}
\label{synchrotron}
We consider a homogeneous and isotropic power-law distribution of relativistic electron energy in the form of 
\begin{equation}
N(E) dE=KE^{2\alpha-1}dE, \label{eq:2}
\end{equation}
where $N$ is number density of relativistic electrons with energy interval between $E$ and $E+dE$, $K$ the normalization factor proportional to electron density, and $\alpha$ the spectral index of electrons.
The synchrotron emission intensity is given by (Ginzburg \& Syrovatskii~\citeyear{Ginzburg1965})
\begin{eqnarray}
I(\nu)=\frac{e^3}{4\pi m_{\rm e} c^2} \int_0^L \frac{\sqrt 3}{2-2\alpha}
\Gamma \left(\frac{2-6\alpha}{12}\right)\Gamma \left(\frac{22-6\alpha}{12}\right)
\nonumber \\
\times \left(\frac{3e}{2\pi m_{\rm e}^3 c^5}\right)^{-\alpha} 
KB_{\perp}^{1-\alpha} \nu^{\alpha}dL, 
\label{eq:I}
\end{eqnarray}
where $\Gamma$ is the Gamma function, $B_{\perp}=\sqrt{B_{x}^2+B_{y}^2}$ the magnetic field component perpendicular to the LOS, and $L$ the integral length along the LOS. The intrinsic synchrotron polarization radiation intensity is calculated by $P_0=Ip$, with the fraction polarization degree being $p=\frac{3-3\alpha}{5-3\alpha}$. The observable Stokes parameters $Q_0$ and $U_0$ are given by $Q_0=P_0\cos 2\psi_{0}$ and $U_0=P_0 \sin 2\psi_{0}$, respectively. 

The synchrotron polarization radiation intensity is depicted as $P\equiv (Q,U)$ in the $Q$-$U$ complex plane, the modulus of which is mathematically written as $\vert P \vert=\sqrt{Q^2+U^2}$. However, the complex modulus $\vert P \vert$ is treated as 
\begin{equation}
P=\sqrt{Q^2+U^2} 
\end{equation}
in astronomy community for simplicity. From a physical point of view, the synchrotron polarization intensity is expressed by 
\begin{equation}
P(\bm X,\lambda^2)=\int_{L_{\rm s1}}^{L_{\rm s2}} dz P_{i}(\bm X,z) e^{2i\lambda^2 {\rm RM}(\bm X,z)} \label{eq:PI}
\end{equation}
along the LOS, where $L_{\rm s1}$ ($L_{\rm s2}$) is lower (upper) boundaries of a synchrotron emitting region. In Equation (\ref{eq:PI}),
$P_{i}\equiv (Q_{i}, U_{i})$ is defined as intrinsic synchrotron polarization intensity {\it density} at a three-dimensional source position ($\bm X$,~z), where the intrinsic polarization information does not suffer from any Faraday rotation effect. On the basis of the synchrotron emission intensity of Equation (\ref{eq:I}) and the subsequent descriptions, we have
\begin{equation}
Q_{i}\propto KB_{\perp}^{-1-\alpha}(B_{x}^2-B_{y}^2)\lambda^{-\alpha}\ \  {\rm and}\  \  U_{i}\propto KB_{\perp}^{-1-\alpha}B_{x}B_{y}\lambda^{-\alpha}, \label{eq:QiUi}
\end{equation}
which are associated with a normalization factor of relativistic electron density $K$, the perpendicular component of magnetic field $B_{\perp}$, and the distribution of wavelength $\lambda^{-\alpha}$.  Since $\lambda^{-\alpha}$ arising from the distribution of relativistic electron cannot result in synchrotron emission fluctuations, this wavelength-dependence included in our simulations cannot change statistical results (LP16 for a theoretical prediction and Zhang, Lazarian \& Xiang~\citeyear{ZhangL18} for numerical confirmation). Thus, synchrotron polarization intensity {\it density} $P_{i}$ at the source can be considered as wavelength-independence, so the wavelength dependence involved in $P(\bm X, \lambda^2)$ is only from Faraday rotation by the factor of $e^{2i\lambda^2 {\rm RM}(\bm X,z)}$.

Considering Faraday rotation effect, the polarization angle of synchrotron radiation is $\psi=\psi_{0}+{\rm RM \lambda^2}$, where $ \lambda$ is the wavelength of synchrotron radiation and RM is the rotation measure. In the case of spatial coincidence, the rotation measure is written as ${\rm RM}=0.81\int_0^z n_{\rm e}B_{\parallel}\,dz~\rm rad~m^{-2}$. In the case of spatial separation, the rotation measure is calculated by 
${\rm RM}=0.81\int_{L_{\rm f1}}^{L_{\rm f2}} n_{\rm e}B_{\parallel}\,dz~\rm rad~m^{-2}$, where $L_{\rm f1}$ and $L_{\rm f2}$ indicate the boundaries of Faraday rotation region. The Faraday rotation density is defined by ${\phi}=n_{\rm e} B_{\parallel}$, where $n_{\rm e}$ is number density of thermal electrons and $B_{\parallel}$ is parallel component of magnetic field along the LOS.

\subsection{Diagnostics of synchrotron polarization}
The ability of gradients of synchrotron polarization intensity $P$ and its wavelength derivative $\frac{dP}{d\lambda^2}$ to trace magnetic field was demonstrated in Lazarian \& Yuen (\citeyear{L2018a}). In addition, Zhang, Liu \& Lazarian~(\citeyear{Zhang19b}) extended the analysis by exploring the tracing ability of other $Q$ and $U$ gradient constructions proposed in Herron et al.~(\citeyear{Herron18a}). Among them, gradients of some synchrotron polarization diagnostics studied are proved to be reliable in tracing magnetic field. Here, we provide several key diagnostics related to this work:\\
1. The magnitude of gradients of complex polarization is
\begin{equation}
{\mathscr P}_{\rm v}=\vert\nabla{P}\vert=\sqrt{(\frac{\partial Q}{\partial x})^2+(\frac{\partial U}{\partial x})^2+(\frac{\partial Q}{\partial y})^2+(\frac{\partial U}{\partial y})^2},
\label{eq:Pv}
\end{equation}
{and its direction is determined by
\begin{eqnarray}
{\rm arg}(\nabla P)={\rm arctan}[{\rm sign}(\frac{\partial Q}{\partial x}
\frac{\partial Q}{\partial y}+\frac{\partial U}{\partial x}\frac{\partial U}
{\partial y})
\nonumber \\
\times \sqrt{(\frac{\partial Q}{\partial y})^2+(\frac{\partial U}{\partial y})^2}/
\sqrt{(\frac{\partial Q}{\partial x})^2+(\frac{\partial U}{\partial x})^2}].
\end{eqnarray}
2. The maximum of the radial component of directional derivative of polarization is
\begin{equation}
{\mathscr P}_{\rm rad}=\sqrt{\frac {(Q \frac{\partial Q}{\partial x}+U \frac{\partial U}{\partial x})^2+(Q \frac{\partial Q}{\partial y}+U \frac{\partial U}{\partial y})^2}{Q^2+U^2}}.
\label{eq:Prad}
\end{equation}
3. The maximum of the tangential component of directional derivative of polarization is
\begin{equation}
{\mathscr P}_{\rm tang}=\sqrt{\frac {(Q \frac{\partial U}{\partial x}-U \frac{\partial Q}{\partial x})^2+(Q \frac{\partial U}{\partial y}-U \frac{\partial Q}{\partial y})^2}{Q^2+U^2}}.
\label{eq:Prad}
\end{equation}
In this paper, derivatives of three quantities $P$, ${\mathscr P}_{\rm rad}$, ${\mathscr P}_{\rm tang}$ are calculated by Sobel operator method, and derivative of ${\mathscr P}_{\rm v}$ is carried out using Equation (\ref{eq:Pv}).
The relative advantages of these gradient diagnostics will be analyzed in Section \ref{sec:results}}. For the rest of the paper, we use Gothic font to denote different methods. 

\subsection{Expectations for the observed signal}

The properties of velocity gradient are the same as those of magnetic field gradient. As is described in Section \ref{sec:theory}, the gradient of magnetic field is perpendicular to magnetic field. Importantly, two recent studies (LP12 and LP16) have explored the direction of magnetic field by fluctuation of synchrotron radiation and synchrotron polarization radiation, from which the direction of magnetic field could be measured by the gradient of fluctuation of them. Analyzing the formulae of polarization intensity correlations, LP16 found that Faraday rotation dominated by mean or random magnetic fields establishes an effective width along the LOS, over which the polarization correlations are accumulated. They proposed a one-radian definition 
\begin{equation}
{\rm RM}{\lambda}^2=0.81{\lambda}^2 \int_0^{L_{\rm eff}}dzn_{e}B_{z}=1
\end{equation}
as the condition for decorrelation of Faraday rotation; this can be generalized to 2$\pi$ radian (see Zhang et al.~\citeyear{Zhang2020}). In view of this, Lazarian \& Yuen (\citeyear{L2018a}) proposed a criterion for collecting effective polarization information, i.e., the ratio of effective polarization region to entire emitting one by ${\frac{L_{\rm eff}}{L}\sim \frac{1}{{\lambda}^2 L}\frac{1}{\phi}}$, where $\phi=\rm max(\sqrt{2}\sigma_{\phi}, \overline{\phi})$ is the maximal value of variance and mean of Faraday rotation measure density and $L$ is the size of emitting region. In the range of the effective length $L_{\rm eff}$, sampled synchrotron radiation is polarized while the radiation is depolarized outside the scale $L_{\rm eff}$, the value of which depends on the wavelength, having smaller (larger) than $L$ for strong (weak) Faraday rotation. 

On the basis of the definition of effective length, observer can actually obtain synchrotron polarization intensity from only a distance smaller than $L_{\rm eff}$. In this regard, the synchrotron polarization intensity is re-written as  
\begin{equation}
P(\bm X, \lambda^2)=\int_{L_{s1}}^{L_{\rm eff}(\lambda)}dz P_{i}({\bf X}, z)e^{2i\lambda^2 {\rm RM}({\bf X},z)}.
\end{equation}
 In general, one can consider the whole foreground region as two parts:  the part of $z<L_{\rm eff}$, which suffers from strong Faraday rotation and the part of $z>L_{\rm eff}$, which does not contribute to synchrotron polarization intensity.

\subsection{{Gradient measurement technique}}
We firstly employ the subblock averaging method (Gonz{\'a}lez-Casanova \& Lazarian~\citeyear{Casanova2017}) to determine the direction of gradients. In each subregion, the optimal direction characterized by the peak of the Gaussian fitting represents the gradient direction. As for the alignment effect between projected magnetic field and gradient direction, we then use the alignment measurement (Gonz{\'a}lez-Casanova \& Lazarian~\citeyear{Casanova2017})
\begin{equation}
\label{eq:am}
AM=\langle2\cos^2\theta-1\rangle,
\end{equation}
to judge their alignment level. In practice, $AM=\pm1$ indicates that the alignment effect is excellent, whereas $AM$=0 represents random orientations. In what follows, we will use $AM$ to quantify the alignment between the gradient of diagnostics (or synchrotron polarization vector) and the projected magnetic field.

\section{Generation of simulation data}
\label{sec:data}

The third-order-accurate hybrid, essentially non-oscillatory code is used to solve control equations of MHD turbulence as follows: 
\begin{equation}
{\partial \rho }/{\partial t} + \nabla \cdot (\rho {\bm v})=0, \label{eq:11}
\end{equation}
\begin{equation}
\rho[\partial {\bm v} /{\partial t} + ({\bm v}\cdot \nabla) {\bm v}] +  \nabla p
        - {\bm J} \times {\bm B}/4\pi ={\bm f}, \label{eq:12}
\end{equation}
\begin{equation}
{\partial {\bm B}}/{\partial t} -\nabla \times ({\bm v} \times{\bm B})=0,\label{eq:13}
\end{equation}
\begin{equation}
\nabla \cdot {\bm B}=0,\label{eq:14}
\end{equation}
where $p=c_{\rm s}^2\rho$ is a gas pressure, $t$ the evolution time of fluid, ${\bm J}=\nabla \times {\bm B}$ the current density, and ${\bm f}$ a random driving force. We simulate a 3D isothermal turbulent medium by considering periodic boundary condition and random solenoidal injection of turbulence driving. The information of data cubes is listed in Table \ref{tab:sim} with a numerical resolution of $512^3$. For the simulation of emitting-source region (see below), the mean magnetic field  $B_0=\langle B \rangle = 1$ is set along the $x$-axis and the LOS is along the $z$-axis. As for the simulation of the foreground region, we change orientations of the mean magnetic field by rotating the data cube.

\begin{table}
\centering
\label{tab:sim}
\begin{tabular}{c c c c c c }
Run & $M_{\rm s}$ & $M_{\rm A}$ & $\beta=2M_{\rm A}^2/M_{\rm s}^2$ & $\delta \rm B_{\rm rms} / \langle B \rangle$   \\ \hline \hline
1 & 9.92 & 0.50 & 0.005 & 0.465   \\
2 & 6.78 & 0.52 & 0.012  & 0.463  \\  
3 & 4.46 & 0.55 & 0.030  & 0.467  \\  
4 & 3.16 & 0.58 & 0.067  & 0.506  \\
5 & 0.87 & 0.70 & 1.295  & 0.579  \\
6 & 0.48 & 0.65 & 3.668   & 0.614  \\
 \hline \hline
\end{tabular}
\caption{Data cubes with numerical resolution of $512^3$ produced by the simulation of sub-Alfv{\'e}nic turbulence. $\delta B_{\rm rms}$ is the root mean square of the random magnetic field of $B$, $\langle B \rangle$ the mean magnetic field, and  $\beta=2M_{\rm A}^2/M_{\rm s}^2$ the plasma parameter.} 
\end{table}

\section{Simulation results}
\label{sec:results}

\begin{figure*}
\centering
\includegraphics[width=0.9\textwidth]{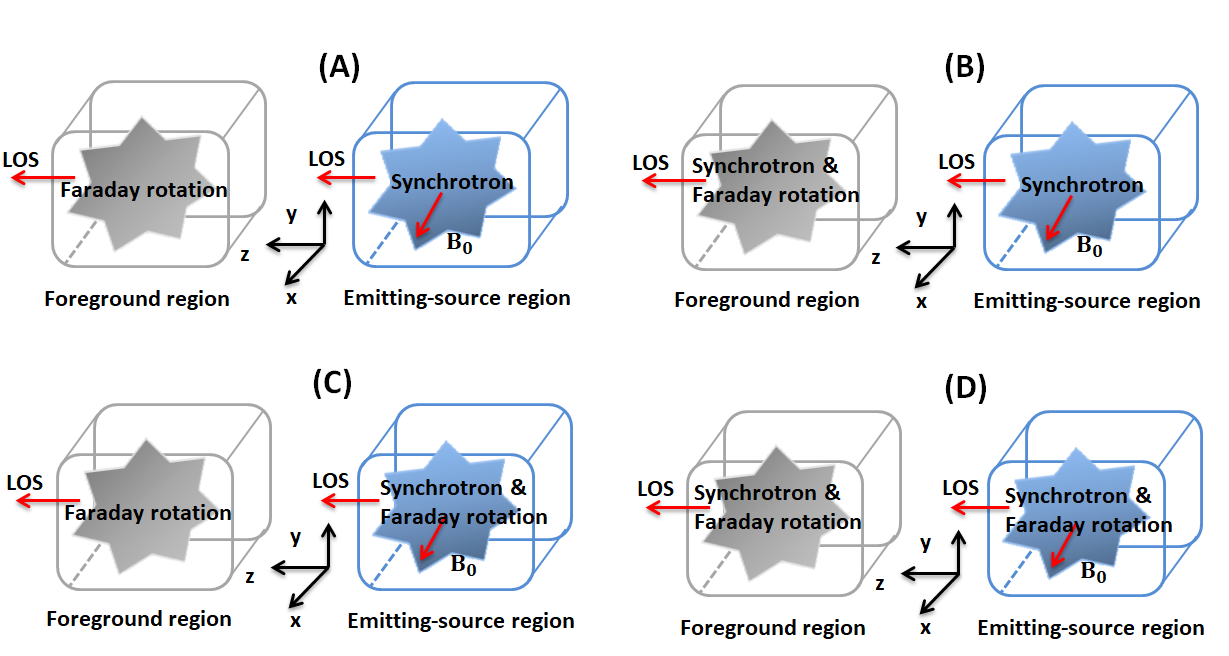}
\caption{ An illustration of spatial configurations for synchrotron emission and Faraday rotation regions. 
The synchrotron polarization radiation results from two spatially separated regions in each case. The mean magnetic field $B_{\rm 0}$ in the emitting-source region is fixed in the $x$-axis direction.}\label{fig:four} 
\end{figure*}

In our previous series of studies, synchrotron gradient technique is only used to explore synchrotron polarization radiation from a single volume, in which polarized synchrotron emission and Faraday rotation are spatially coincident. In this paper, we consider spatially separated situation for synchrotron polarization and Faraday rotation regions (see Figure \ref{fig:four} with four cases). Each case includes two parts, the right part of which is called emitting-source region, and the left one is the foreground region. In our calculation, thermal electron densities are set at $1.0 ~\rm cm^{-3}$ for the former and $0.1 ~\rm cm^{-3}$ for the latter, respectively. We consider the two parts different magnetic field strengths, i.e., 1.0 $\mu G$ for emitting-source region and 0.8 $\mu G$  for foreground region. Additionally, we assume that the two regions have the same spatial scale 50 $\rm pc$ along the LOS. In this section, we discuss the mean magnetic field within foreground region along the $y$-axis.

\begin{figure*}
\centering
\includegraphics[width=0.95\textwidth]{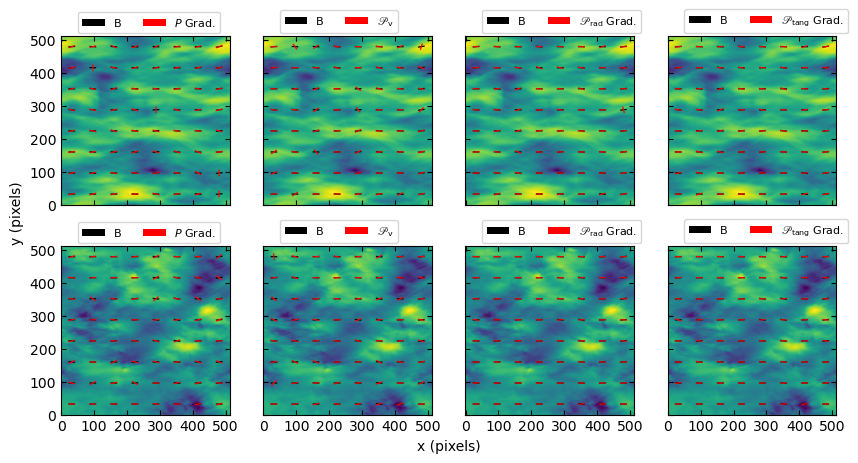}
\caption{A comparison between the directions of projected magnetic field in the emitting-source region and the directions predicted by the gradient measurement of various diagnostics for Case A (see Figure \ref{fig:four}). The background map in a logarithmic scale is an image of synchrotron polarization intensity in units of its mean value. The upper and lower panels represent subsonic (see run6 in Table \ref{tab:sim}) and supersonic (see {run3} in Table \ref{tab:sim}) turbulence, respectively. The alignment measures from gradient of $P$ and ${\mathscr P}_{\rm v}$ are AM$\approx 0.7$ for the former and AM$\approx 0.8$ for the latter. The alignment measurements from gradients of both ${\mathscr P}_{\rm rad}$ and ${\mathscr P}_{\rm tang}$ are approximately 0.9.}\label{fig:four_map} 
\end{figure*}

Using alignment measurement, we test whether the gradient of synchrotron polarization $P$ or any of three gradient diagnostics (${\mathscr P}_{\rm v}$, ${\mathscr P}_{\rm rad}$, ${\mathscr P}_{\rm tang}$) mentioned in Section \ref{synchrotron} is better for tracing projected magnetic field. With the setting of $\alpha=-1$ and $\nu=0.1~\rm GHz$, Figure~\ref{fig:four_map} shows the distributions of alignment measurements arising from subsonic (upper panels) and supersonic (lower panels) turbulence. It is shown that the gradients of diagnostic techniques (${\mathscr P}_{\rm rad}$, ${\mathscr P}_{\rm tang}$) have a better alignment measurement compared with other diagnostic techniques. In view of our findings in Zhang, Liu \& Lazarian~(\citeyear{Zhang19b}) that the gradient of ${\mathscr P}_{\rm rad}$ has a robust ability for tracing magnetic field, ${\mathscr P}_{\rm rad}$ will be chosen to trace projected magnetic field in the following studies. Throughout this paper, we use a Gaussian kernel of less than 2$\sigma$ to smooth small scale noise-like structures. 

\begin{figure*}
\centering
\includegraphics[width=0.9\textwidth]{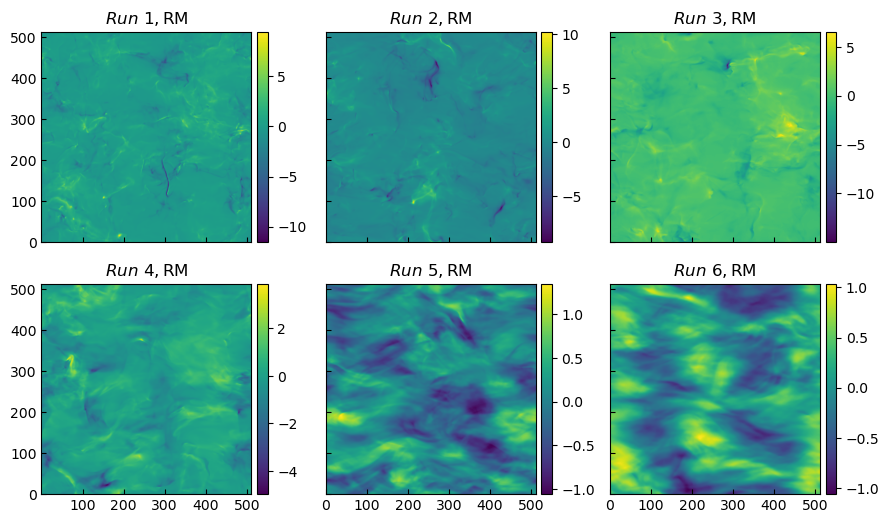}
\caption{The 2D structure map of Faraday measure in units of rad m$^{-2}$ on the basis of data listed in Table \ref{tab:sim}  .}\label{fig:fr_six} 
\end{figure*}

Before studying the alignment measurements in four spatial configurations, we first explore the magnitude of rotation measure in foreground region. Figure \ref{fig:fr_six} displays the 2D structure map of rotation measure integrated along the LOS. It is shown that rotation measure for subsonic turbulence is smaller than that for supersonic turbulence. The reason lies in that supersonic turbulence, whose high density region leads to a large $\phi$ value due to $\phi \propto n_{\rm e}B_{\parallel}$, has more significant density inhomogeneity.

\subsection{Synchrotron polarization emitting-source without Faraday rotation effect}

\begin{figure}
%\centering
\includegraphics[width=0.5\textwidth]{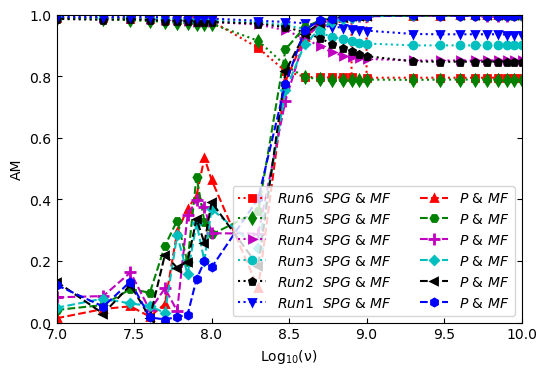}
\caption{ Alignment measure as a function of radio frequency for Case A (see Figure \ref{fig:four}). The legend RunX SPG \& MF represents the alignment measure of the gradient of ${\mathscr P}_{\rm rad}$ and projected magnetic fields in the emitting-source region. The legend P \& MF on the right hand indicates the corresponding alignment measure of the 90-degree-rotated synchrotron polarization vector and the projected magnetic field.}\label{fig:caseA} 
\end{figure}

\begin{figure}
%\centering
\includegraphics[width=0.5\textwidth]{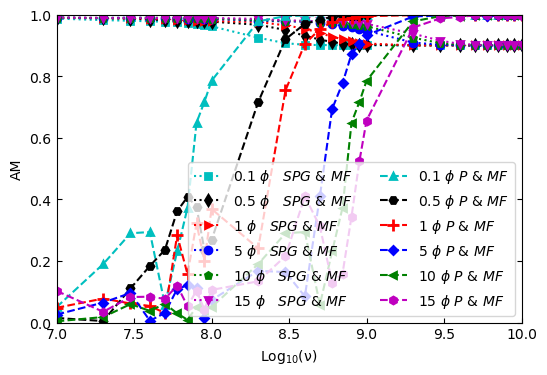}
\caption{ Alignment measurement as a function of radio frequency for the simulation of $M_{\rm s}= 4.46$ and $M_{\rm A}=0.55$ for Case A. The legend $x$~$\phi$ represents $x$ times Faraday rotation density in the foreground region.
}\label{fig:caseAFR} 
\end{figure}

\subsubsection{Case A: Faraday rotation effect in foreground region}

The first to explore is Case A (see Figure \ref{fig:four} (A)), that is, synchrotron polarization radiation from emitting-source region is subject to Faraday rotation in foreground region. We investigate AMs of gradient of ${\mathscr P}_{\rm rad}$ and polarization versus projected magnetic field in a broadband frequency range (0.01 to $10 ~\rm GHz$) using simulation data listed in Table \ref{tab:sim}.

\begin{figure}
%\centering
\includegraphics[width=0.5\textwidth]{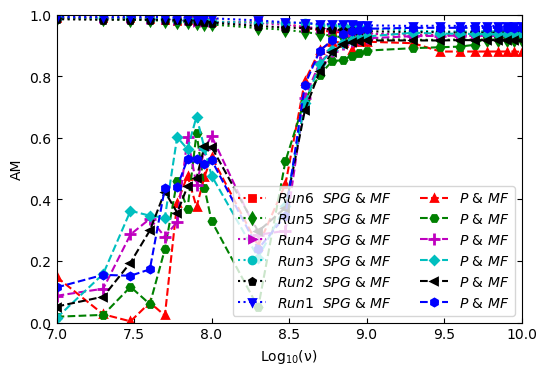}
\caption{ Alignment measurement for Case B. The other descriptions are the same as those of Figure \ref{fig:caseA}. }\label{fig:caseB} 
\end{figure}

Figure \ref{fig:caseA} represents alignment measurement of gradient of ${\mathscr  P}_{\rm rad}$ and projected magnetic field, compared with traditional polarization vector method. It is clearly seen that large AM values from the gradient measurement span over a wide low frequency range, while slightly decreasing AM values appear in the high frequency regime. Based on tomographic studies of Faraday depolarization, we know that correlating synchrotron polarization radiation is fully sampled from a small part of the entire foreground region in the low frequency range. Thus, noise-like structures in the remaining part should not affect the alignment measurement due to a Gaussian filter. In contrast, polarized radiation correlations are collected beyond the whole foreground region in the high frequency range. This insufficient sampling could lead to a decrease in AM. In addition, numerical resolution of data could influence the alignment measurements. We find that the projected magnetic field traced by gradient techniques is not subject to Faraday rotation, which is consistent with the situation for spatially coincident synchrotron polarization radiation and Faraday rotation (Zhang et al.~\citeyear{Zhang19a}; Zhang, Liu \& Lazarian~\citeyear{Zhang19b}).

The AMs obtained by polarization vector method are small (about zero) in the low frequency range but larger (about 1) in high the frequency regime. 
The failure of polarization vector method in the low frequency range is due to strong Faraday rotation effect from foreground region that impedes from measurement of projected magnetic field within the emitting-source region, but Faraday rotation effect becomes weak in the high frequency range.
Therefore, magnetic field traced by gradient techniques is not subject to Faraday rotation, which is consistent with the situation for spatially coincident synchrotron polarization radiation and Faraday rotation (Zhang et al.~\citeyear{Zhang19a}; Zhang, Liu \& Lazarian~\citeyear{Zhang19b}).
Here, polarization radiation in different spatial locations suffers from differential Faraday rotation effects along the LOS. It should be emphasized that the currently separated scenario in space confronts with more Faraday rotation effect than spatially coincident one, because synchrotron polarization emissions arising from emitting-source region undergo the same high level of Faraday rotation effects when they propagate through the entire foreground region. 
 
\begin{figure}
%\centering
\includegraphics[width=0.5\textwidth]{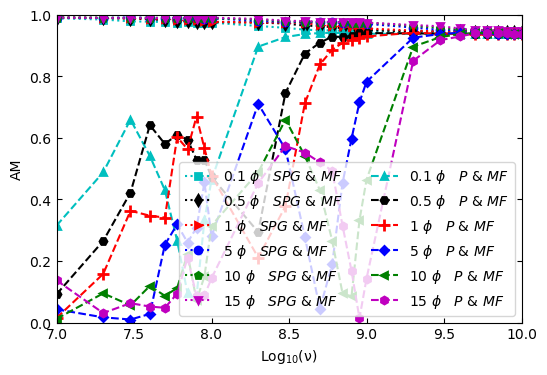}
\caption{ Alignment measurement for Case B based on Run3. The other descriptions are the same as those of Figure \ref{fig:caseAFR}. }\label{fig:caseBFR} 
\end{figure}

To explore the influence of Faraday rotation effect from foreground region on measurement of projected magnetic field for Case A, we calculate AMs in Figure \ref{fig:caseAFR} based on Run3 by varying Faraday rotation density. As shown in Figure \ref{fig:caseAFR}, the AMs between gradient of ${\mathscr P}_{\rm rad}$ and projected magnetic field are slightly affected in the broadband frequency range by different Faraday rotation densities, i.e., different levels of Faraday depolarization. With the increase of Faraday rotation density, the distributions of AMs between synchrotron polarization vector and projected magnetic field shift to higher frequency regime. The traditional method of tracing magnetic field, i.e., synchrotron polarization vector, is subject to Faraday rotation effect and cannot work in the low frequency range. Interestingly, the gradients of ${\mathscr P}_{\rm rad}$ insensitive to the level of Faraday rotation can provide an excellent opportunity for measuring the magnetic field directions.

\subsubsection{Case B: Faraday rotation effect accompanied by polarization radiation in foreground region}

In this section, we consider that synchrotron polarization radiation from emitting-source and foreground regions is affected by Faraday rotation in  foreground region (see Case B of Figure \ref{fig:four}). Figure \ref{fig:caseB} shows how radio frequency affects gradient technique measurement for projected magnetic field within emitting-source region, compared with traditional polarization vector method. As shown in the dashed lines, AMs between polarization vector and projected magnetic field are close to zero in the low frequency from 0.01 to $0.1 ~\rm GHz$, distribution of which presents bumps at the frequency $0.1 ~\rm GHz$. 
Additionally, the distribution of AMs also displays an obvious trough at the frequency $0.2 ~\rm GHz$. In this case, the appearance of bumps may imply that polarized radiation at different depths along the LOS experiences weaker Faraday rotation effect, while the trough corresponds an opposite scenario. We find that gradient techniques can well trace magnetic field direction through the entire frequency range. 

Similarly, we also consider the effect of Faraday rotation density on alignment measurement at different frequencies, based on simulation of Run3. As shown in Figure \ref{fig:caseBFR}, large AMs between gradient of ${\mathscr P}_{\rm rad}$ and projected magnetic field are almost constant at all the frequencies, which are insensitive to Faraday rotation density. With increasing Faraday rotation density, the small AM distributions between synchrotron polarization vector and projected magnetic field move up to higher frequency regime, implying the change in differential Faraday rotation depolarization. 

\subsection{Synchrotron polarization emitting-source with Faraday rotation effect}

\subsubsection{Case C: Faraday rotation effect in foreground region}
\begin{figure}
%\centering
\includegraphics[width=0.5\textwidth]{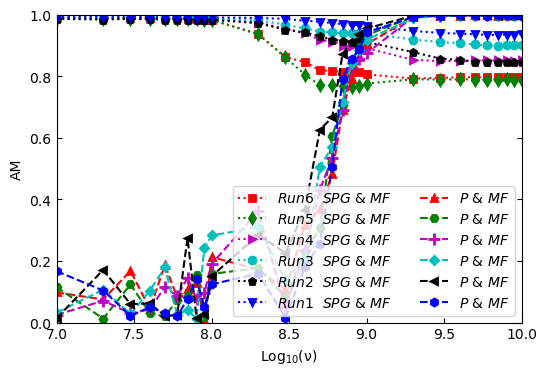}
\caption{ Alignment measurement at different frequencies for Case C. }\label{fig:caseC} 
\end{figure}

\begin{figure}
%\centering
\includegraphics[width=0.5\textwidth]{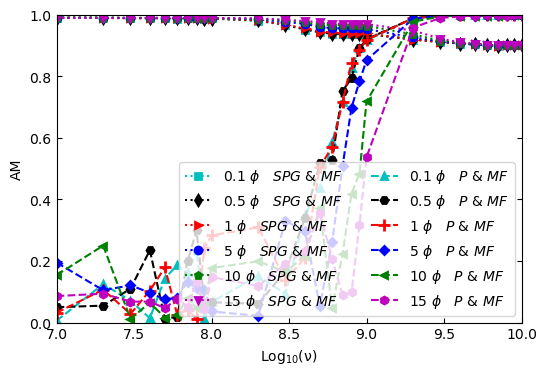}
\caption{ Alignment measurement at different Faraday rotation densities for Case C. }\label{fig:caseCFR} 
\end{figure}

Based on the spatial configuration in Case C, we study the measurement of magnetic field direction using polarization gradient techniques. Specifically, we consider how different turbulence types listed in Table \ref{tab:sim} and Faraday rotation density affect AMs between gradient of ${\mathscr P}_{\rm rad}$ and projected magnetic field. Figure \ref{fig:caseC} shows that the trend of AMs of Case C obtained by gradient technique is almost consistent with Case A. However, there is a certain difference of AMs in polarization vector method, that is, the distribution of AM values close to 1 for Case C locates at higher frequencies. In other words, Case C experienced stronger Faraday rotation effect than Case A, which is in line with expectation. Furthermore, Figure \ref{fig:caseCFR} plots the influence of Faraday rotation density on alignment measurement. It can be seen that the distribution of AMs obtained by gradient technique is almost the same when changing Faraday rotation density, which implies the AM change independent of Faraday rotation density. However, the AM values in this case by polarized vector method increase at the frequency of $0.3~\rm GHz$, which is relatively lagging compared with those of Case A, indicating a stronger Faraday rotation density.

\subsubsection{Case D: Faraday rotation effect accompanied by polarization radiation in foreground region}
\begin{figure}
%\centering
\includegraphics[width=0.5\textwidth]{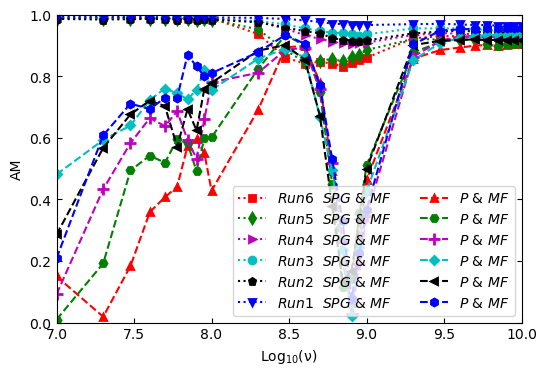}
\caption{ Alignment measurement at different frequencies for Case D.} 
\label{fig:caseD}
\end{figure}

\begin{figure}
%\centering
\includegraphics[width=0.5\textwidth]{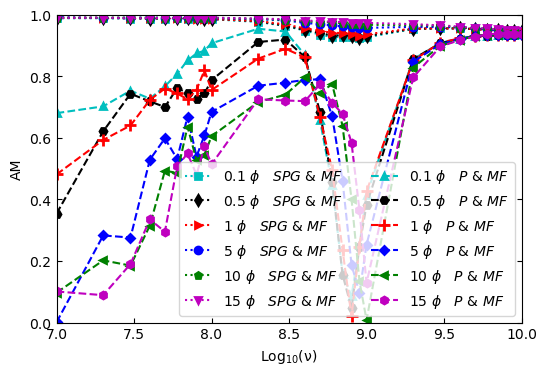}
\caption{ Alignment measurement at different Faraday rotation densities for Case D.}\label{fig:caseDFR} 
\end{figure}

The numerical results are plotted in Figures \ref{fig:caseD} and \ref{fig:caseDFR} for Case D. In general, large AM values demonstrate that synchrotron polarization gradient techniques can provide a good way for measuring projected magnetic field directions. The dashed lines in Figures \ref{fig:caseD} and \ref{fig:caseDFR} represent AMs between synchrotron polarization vector and projected magnetic field within synchrotron emitting source. Figure \ref{fig:caseD} shows that the distribution of AM values obtained by polarized vector method has a roughly N-like distribution, i.e., indicating the instability of measurement of polarization vector method. It can be seen that the trend of AM values for Case D is similar to that of Case B, but its magnitude is different. The larger AM value represents weak Faraday rotation effect. As a result, synchrotron polarization gradient techniques can work well for measuring magnetic field in various spatial configurations. In particular, this technique is suitable for exploring magnetic field properties under a strong Faraday rotation condition.

\subsection{Influence of mean magnetic field orientations within foreground region on AMs}

\begin{figure*}
\centering
\includegraphics[width=0.9\textwidth]{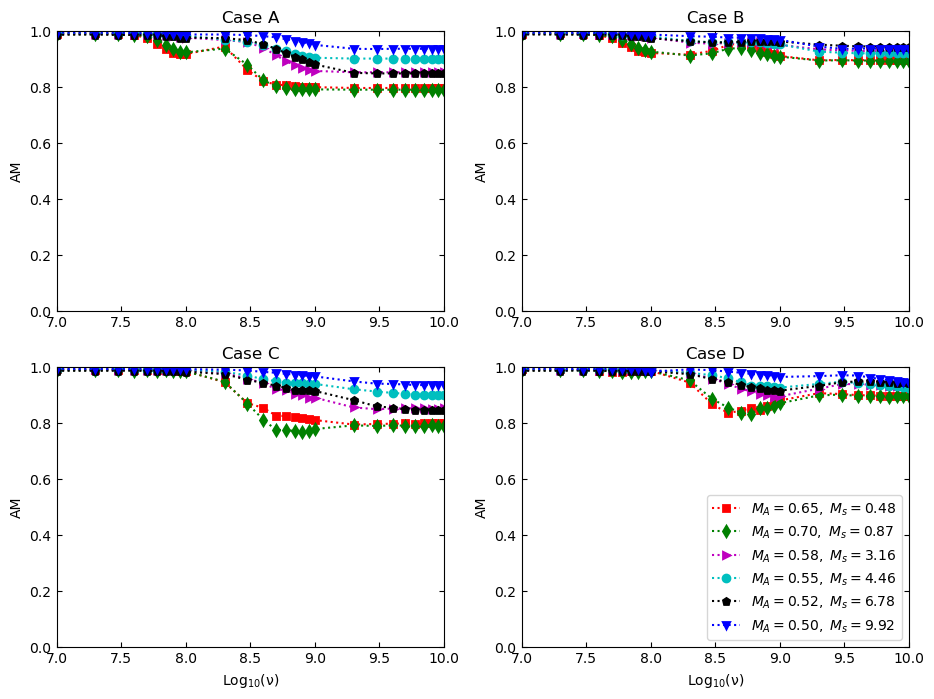}
\caption{The AMs between the gradient of ${\mathscr P}_{\rm rad}$ and projected magnetic field for four spatial configurations. The angle between mean magnetic field of foreground region and the $x$-axis is 60 degrees.
}\label{fig:rotation60} 
\end{figure*}

In the previous sections, the mean magnetic field in the foreground region is set along $y$-axis as an example. Keeping the same LOS direction and emitting-source region as shown in Figure \ref{fig:four}, we explore how the angle between mean magnetic field of foreground region and the $x$-axis affects the alignment measurement. The results for 60 degrees plotted in Figure \ref{fig:rotation60} demonstrate that synchrotron polarization gradient technique still provides the reliability of the measurement for projected magnetic field direction. In addition, we also find that the results in the case of 30 degrees show a great similarity to those of 60 degrees.

\begin{figure}
%\centering
\includegraphics[width=0.5\textwidth]{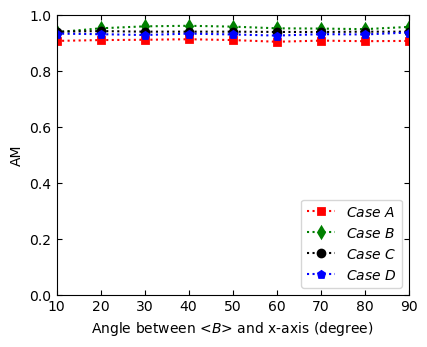}
\caption{ Alignment measurement as a function of the angle between mean magnetic field of foreground region and the $x$-axis for four configurations. }\label{fig:angle} 
\end{figure}

In case of certain limitations when confronting with complex astrophysical environment, we hereby fix frequency at 1 GHz and study AM distribution as a function of the angle. As shown in Figure \ref{fig:angle}, the AMs remain large values as expected at different angles for four cases by using the gradient of ${\mathscr P}_{\rm rad}$. Although various angles between mean magnetic field of foreground region and the $x$-axis do not affect Faraday rotation density, they can change the purely polarized emission information in the local foreground region (see Cases B and D of Figure \ref{fig:four}). 

Meanwhile, setting the mean magnetic field $B_0$ in the foreground region along the LOS, we explore the alignment measurements for four spatial configurations shown in Figure \ref{fig:four}. We find that the AMs obtained by the gradient technique of ${\mathscr P}_{\rm rad}$ are still large enough, up to AM$\simeq$0.9. Evidently, our gradient measurements demonstrate that the complex and changeable magnetic field configurations within the foreground region do not affect the projected magnetic field measurement from the emitting-source regions.

\section{Conclusions and Discussion}
\label{sec:discussions}

This work has studied the ability of SGT to trace magnetic fields in diffuse interstellar medium. In particular, we applied synchrotron polarization gradient technique to spatially separated configurations in order to simulate more complex ISM environment. We proved that the combinations of Stokes parameters $Q$ and $U$  employed for gradient studies of synchrotron polarization are working well. We mainly focused on using gradient of ${\mathscr P}_{\rm rad}$ to recover projected magnetic field directions within emitting-source region at multi-frequency bands, in contrast with traditional polarization vector method. We find that for various spatially separated polarization emission and Faraday rotation regions, gradient techniques of diagnostic ${\mathscr P}_{\rm rad}$ perform very well in measuring projected magnetic field, mostly independent of radio frequency. At the same time, the polarization vector method itself fails to trace magnetic field direction in the presence of Faraday rotation. 

Since AMs between gradient of ${\mathscr P}_{\rm rad}$ and projected magnetic field are more than 0.9 on the basis of numerical results, the gradient of ${\mathscr P}_{\rm rad}$ could probe well projected magnetic field within emitting-source region at all the frequency bands studied, although gradient techniques cannot identify  
whether observational signal of synchrotron polarization radiation is from the emitting-source or foreground region. As for traditional polarization vector method, which cannot measure magnetic field properties in the low-frequency, strong Faraday rotation regime, it seems able to distinguish frequency-dependent depolarization or differential Faraday rotation depolarization. In the case of weak Faraday rotation depolarization, two methods are synergetic for studying the properties of MHD turbulence in spatially separated situations.

Our numerical results show that gradient of ${\mathscr P}_{\rm rad}$ could well trace projected magnetic field direction through almost entire frequency range with an exception of slightly decreasing AMs at high frequency end. Therefore, we would like to claim that ${\mathscr P}_{\rm rad}$ is a better probe for tracing magnetic field properties. In the case of subsonic turbulence, i.e, Case A, Figure \ref{fig:four_map} showed that  ${\mathscr P}_{\rm v}$ and the gradient of $P$ could not trace well projected magnetic field in emitting-source region, which highlights the need for multiple synergetic techniques when studying complex astrophysical environments. It should be emphasized that in the emitting-source region, the direction of mean magnetic field $B_0$ was considered along $x$-axis. As for the foreground region, $B_0$ can be changeable, i.e., arbitrary in the $x$-$y$ plane or along the $z$ direction.

Correlation of synchrotron polarization radiation can be sampled over the effective width in the low frequency range, but not fully collected in the high frequency range which results in a slightly decreased AM. Besides, limited numerical resolution in current work may also account for the reduced AM at high frequency regime, compared with Figure 6 of  Zhang, Liu \& Lazarian~(\citeyear{Zhang19b}). Compared with synchrotron intensity gradients (Lazarian et al.~\citeyear{L2017}) independent of wavelength and free from the Faraday rotation effect, Faraday depolarization, a new way to restore 3D magnetic field structures, is the most promising effect related to synchrotron polarization gradients. The synergetic usage of the two techniques can improve the reliability of tracing magnetic field.

Our earlier studies found that the anisotropy of polarized synchrotron intensity could be used to estimate the direction of projected magnetic field roughly. This is still a complementary method to synchrotron gradient techniques that can trace projected magnetic field directions more accurately. The advantage of structure functions of synchrotron and polarization intensities is that their statistical methods have been described in detail in LP12 and LP16, which establishes a theoretical basis of synchrotron statistical analysis and provides possibilities to distinguish contribution of three basic MHD modes, i.e. Alfv{\'e}n, slow and fast modes. The spatially separated configurations were first proposed in Zhang, Lazarian \& Xiang~(\citeyear{ZhangL18}) to simulate complex ISM for recovering the spectral properties of MHD turbulence at different wavelengths. Similarly, the anisotropy of polarized synchrotron intensity at different wavelengths was studied in Lee, Cho \& Lazarian~(\citeyear{Lee2019}) using structure function and quadrupole moment. The purpose of the current work is furthering new gradient techniques for more realistic astrophysical scenario. The accumulation of large amounts of observational data facilitates the application of new magnetic field measurement techniques. For instance, the gradient techniques proposed in our series studies have been applied to realistic observational data from Plank (Yuen \& Lazarian~\citeyear{Yuen2017}) and Canadian Galactic Plane Survey (Zhang, Liu \& Lazarian~\citeyear{Zhang19b}). Recently, observations from Urumqi 6 cm polarization survey has been used to identify different plasma modes in the Galactic turbulence (Zhang et al~\citeyear{zhang2020nature}).

Despite the synchrotron polarization gradients as the focus in this paper to trace the projected magnetic field directions, the distribution of gradients over the sub-block can also obtain magnetization of media by Alfv{\'e}nic Mach number $M_A$ (Lazarian \& Yuen \citeyear{L2018b}, Carmo et al. \citeyear {Carmo2020}). The value of $M_A$ is instrumental for describing many key astrophysical processes including the star formation and cosmic ray propagation. In addition, the value of $M_A$ can be used to obtain magnetic field using the technique described in Lazarian, Yuen \& Pogosyan~(\citeyear{Lazarian2020b}). We plan to address the issue of obtaining $M_A$ using synchrotron gradients elsewhere.

\section*{ACKNOWLEDGMENTS}
J.F.Z. acknowledges the support from the National Natural Science Foundation of China (grants No. 11973035 and 11703020), the Hunan Province Innovation Platform and Talent Plan--HuXiang Youth Talent Project (No. 2020RC3045). A.L. thanks the support of NSF AST 1816234 and NASA TCAN  144AAG1967 and NASA AAH7546 grants. F.Y.X. acknowledges the support from the Joint Research Funds in Astronomy U2031114 under cooperative agreement between the National Natural Science Foundation of China and the Chinese Academy of Sciences.

\section*{DATA AVAILABILITY}
The data underlying this paper can be shared on reasonable request to the corresponding author.

% Don't change these lines
\bsp    % typesetting comment
\label{lastpage}

\end{document}